\documentstyle[aas2pp4]{article}

\lefthead{Shirey et al.}
\righthead{Quasi-periodic Oscillations in Cir X-1}

\newcommand{\RXTE}{{\it RXTE\,}}
\newcommand{\EXOSAT}{{\it EXOSAT\,}}
\newcommand{\ASCA}{{\it ASCA\,}}
\newcommand{\Ginga}{{\it Ginga\,}}
\newcommand{\intens}[2]{I\,[#1--#2~keV]}
\newcommand{\softcolor}{\intens{4.8}{6.3}~/ \intens{2.0}{4.8}}
\newcommand{\broadcolor}{\intens{6.3}{13}~/ \intens{2.0}{6.3}}
\newcommand{\hardcolor}{\intens{13}{18}~/ \intens{8.5}{13}}

\begin{document}

\title{Quasi-periodic Oscillations Associated with Spectral Branches 
in \RXTE\ Observations of Circinus X-1}
\author{Robert E. Shirey, Hale V. Bradt, Alan M. Levine, \& Edward H. Morgan}


\affil{
Department of Physics and Center for Space Research, \\
Massachusetts Institute of Technology, Cambridge, MA 02139; \\
shirey@space.mit.edu 
\\ 
{\bf To be published in October 10, 1998 issue of The Astrophysical Journal (Vol. 506)}
} 

\begin{abstract}
We present {\it Rossi X-ray Timing Explorer}\ (\RXTE) All-Sky Monitor
observations of the X-ray binary Circinus~X-1 which illustrate the
variety of intensity profiles associated with the 16.55~d flaring
cycle of the source. We also present eight observations of Cir~X-1
made with the \RXTE\ Proportional Counter Array over the course of a
cycle wherein the average intensity of the flaring state decreased
gradually over $\sim$12 days. Fourier power density spectra for these
observations show a narrow quasi-periodic oscillation (QPO) peak which
shifts in frequency between 6.8~Hz and 32~Hz, as well as a broad QPO
peak that remains roughly stationary at $\sim$4~Hz. We identify these
as Z-source horizontal and normal branch oscillations (HBOs/NBOs)
respectively. Color-color and hardness-intensity diagrams (CDs/HIDs)
show curvilinear tracks for each of the observations. The properties
of the QPOs and very low frequency noise allow us to identify segments
of these tracks with Z-source horizontal, normal, and flaring branches
which shift location in the CDs and HIDs over the course of the
16.55~d cycle. These results contradict a previous prediction, based
on the hypothesis that Cir~X-1 is a high-$\dot{M}$ atoll source, that
HBOs should never occur in this source (\cite{oosterbroek95};
\cite{klis94}).

\end{abstract}

\keywords{Stars:individual(Cir~X-1) --- stars:neutron --- X-rays:stars}

\section{Introduction}

The X-ray binary Circinus~X-1 is unique in its complex temporal and
spectral variability. A 16.55~day cycle of flaring is observed in the
X-ray (\cite{kaluzienski76}) as well as optical (\cite{moneti92}), IR
(\cite{glass78}), and radio bands (\cite{whelan77}). The high degree
of stability of the period of this cycle is evidence that it is
the orbital period.  The onset of flaring has been suggested to be the
result of enhanced mass transfer occurring near periastron of a
highly eccentric binary orbit (\cite{murdin80}; \cite{oosterbroek95}).

The X-ray profile and average intensity of the 16.55~day cycle has
varied considerably over timescales of years (see e.g, Dower, Bradt,
\& Morgan 1982\nocite{dower82}; \cite{stewart91};
\cite{oosterbroek95}; Shirey et~al.\ 1996, hereafter
\cite{shirey96}). Observations with the {\it Rossi X-ray Timing
Explorer} (\RXTE) All-Sky Monitor (ASM, 2--12~keV) showed Cir~X-1 in a
sustained bright state with a baseline intensity level of
$\sim$1.0~Crab (75 c/s; 1060 $\mu$Jy at 5.2 keV) and strong flaring up
to as high as 3.5~Crab (\cite{shirey96}). The flaring state began
during the day following phase zero (based on the radio ephemeris of
\cite{stewart91}) and typically lasted 2--5~days (see
Figure~\ref{fig:asm} and discussion below). The ratio of count rates
in different ASM energy channels showed dramatic spectral softening at
the onset of the flaring state and gradual hardening during the
remainder of each cycle (\cite{shirey96}). Similar behavior was seen
in \Ginga\ ASM observations folded at the 16.55 day period
(\cite{tsunemi89}).
Near phase zero, some of the cycles observed with the \RXTE\ ASM also
showed brief dips below the 1~Crab level. Observations of dips near
phase zero in Cir~X-1 with \ASCA\ (\cite{brandt96}) and the \RXTE\ PCA
(\cite{bradt98}) indicate the presence of both a strongly absorbed
spectral component and an unabsorbed component.

Observations of type~1 X-ray bursts demonstrate that Cir~X-1 is a
low magnetic field neutron star (Tennant, Fabian, \& Shafer
1986\nocite{tennant86b}). Additional type~1 bursts have not been
observed from Cir~X-1 since the \EXOSAT\ discovery, possibly because
the source intensity has been higher during subsequent observations.

The rapid X-ray variability of Cir~X-1 at times resembles that of both
``atoll'' and ``Z'' low-mass X-ray binaries (LMXBs) as well as
black-hole candidates (\cite{oosterbroek95}). Quasi-periodic
oscillations (QPOs) were reported at 1.4~Hz, 5--20~Hz, and 100--200~Hz
in \EXOSAT\ data (\cite{tennant87}, 1988\nocite{tennant88}). Based on
these data, it has been suggested that Cir~X-1 is an atoll source that
can uniquely reach the Eddington accretion rate and exhibit
normal/flaring branch QPOs at 5--20~Hz (\cite{oosterbroek95};
\cite{klis94}). Observations made during non-flaring phases with the
\RXTE\ Proportional Counter Array (PCA) showed a QPO peak that varied
from 1.3 to 12~Hz, flat-topped low-frequency noise (LFN), and a broad
peak that varied from 20--100~Hz (\cite{shirey96}). The two QPO
frequencies and the cut-off frequency of the flat-topped noise were
highly correlated. These QPOs are likely to be essentially the same
phenomenon as those previously seen in the \EXOSAT\ observations at
5--20~Hz and 100--200~Hz.

In this paper we present additional \RXTE\ ASM observations of Cir~X-1
which further illustrate how its intensity profile varies from one
16.55~d cycle to another. We also present the results of \RXTE\ PCA
observations made over the course of one cycle in which the intensity
declined unusually gradually from the flaring state to the quiescent
level. This slow transition allows us to demonstrate how the
time-variability properties of the source are related to its spectral
properties.

\section{Observations}

\begin{figure}
\epsscale{1}
\plotone{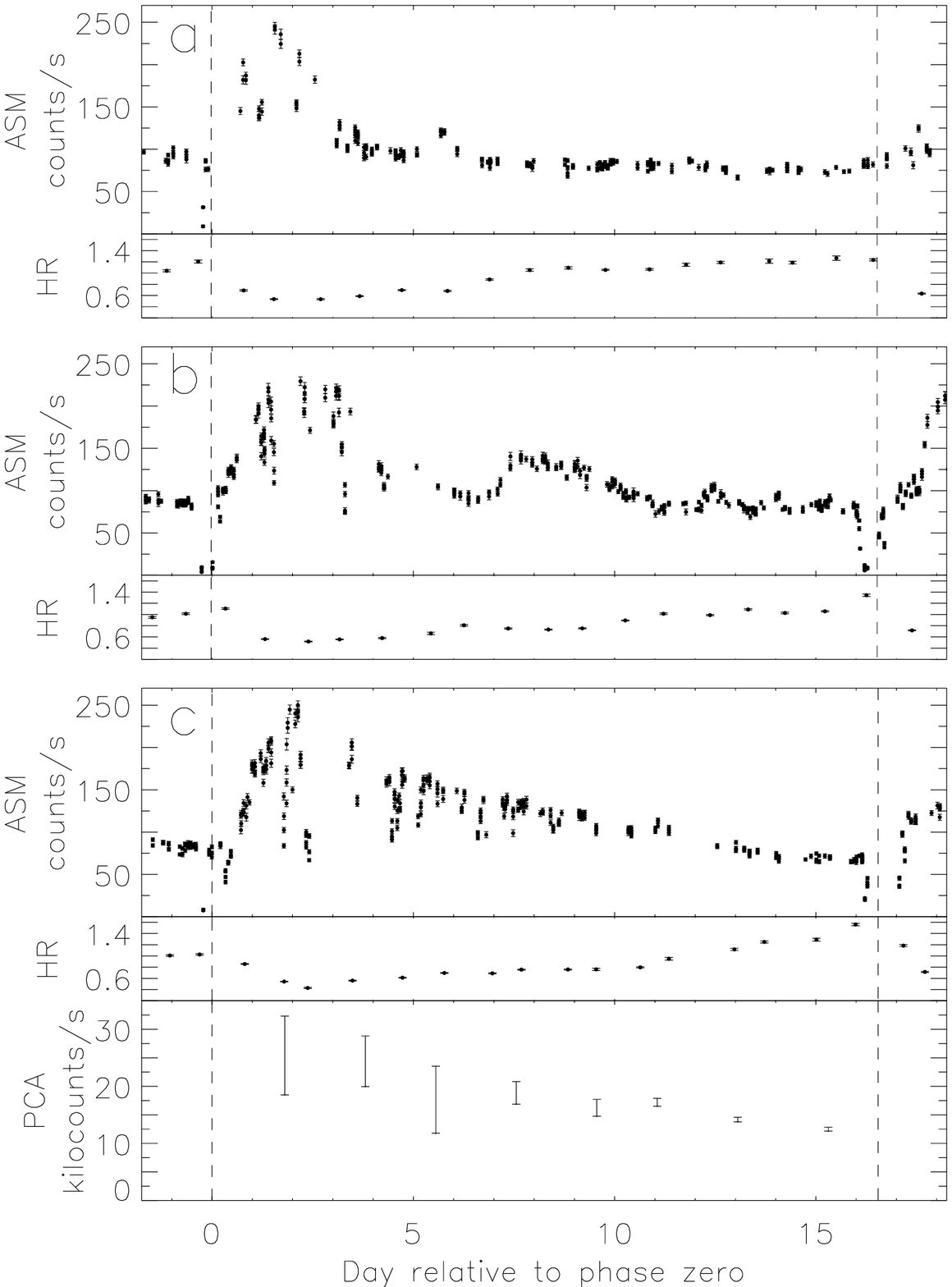}
\caption{ 
\RXTE\ ASM light curves (1.5--12~keV) for three 16.55~d cycles of
Cir~X-1 showing different flaring profiles. Each intensity point
corresponds to a 90-s exposure by one of the three ASM cameras, and
the hardness ratio (HR), defined as the ratio of counting rates for
5--12~keV to 3--5~keV, is shown in one-day averages. The 3--5~keV to
1.5--3~keV hardness ratio exhibits very similar behavior and is not
shown here. The intensities are for Cir~X-1 after background and other
sources in the field of view have been subtracted. The Crab nebula
yields $\sim$75~c/s.  Vertical dashed lines indicate phase zero based
on the radio ephemeris of Stewart et~al.\
(1991\nocite{stewart91}). Day zero corresponds to (a) 1997 April~23.87
(b) 1996 August~2.14 and (c) 1997 February~16.69. For cycle (c), the
intensity ranges (\intens{2.0}{18}) seen in the eight \RXTE\ PCA
observations (I-VIII in time order) are also shown.
\label{fig:asm}}
\end{figure}

The \RXTE\ ASM has now provided 2--12~keV light curves for over 45
orbital cycles of Cir~X-1 since 1996 February. Throughout all these
cycles, the baseline intensity has remained near 1.0~Crab. The variety
of intensity profiles is illustrated in Figure~\ref{fig:asm}, which
shows ASM light curves and hardness ratios for three cycles.  In many
cycles, after 3--5~days in the flaring state the intensity is quite
steady for the remainder of the cycle (e.g., Figure~\ref{fig:asm}a).
In addition to the main flaring episode, some cycles show a mid-phase
flare (not always at the same phase) to as high as 2~Crab
(Figure~\ref{fig:asm}b). Occasionally, the flaring state begins after
phase zero and continues for most of the cycle with a gradually
decreasing intensity (Figure~\ref{fig:asm}c).
All cycles observed with the ASM show the general pattern of spectral
hardening mentioned above. During the half day before phase zero, and
continuing intermittently for up to two days, brief dips occur in many
cycles (perhaps in all cycles, since the ASM coverage is incomplete).
These dips are seen as isolated low points in the ASM light curves of
Figure~\ref{fig:asm}.

Eight PCA observations ($\sim$6~ksec each) were carried out during
1997 February~18 -- March~4 to sample one 16.55~day cycle at roughly
two-day intervals (Figure~\ref{fig:asm}c and Table~\ref{tab:obs}). The
very gradual decline of the flaring-state intensity in this cycle
serendipitously provided an opportunity to study intensity-related
source properties. All five proportional counter units (PCUs) of the
PCA operated normally during each observation, except during the first
few minutes of the first observation when only three PCUs were on. All
intensities for that period have been adjusted by a factor of 5/3, but
these data are not used in color-color and hardness-intensity diagrams
due to gain differences between detectors.

\begin{figure*}
\epsscale{1.8}
\plotone{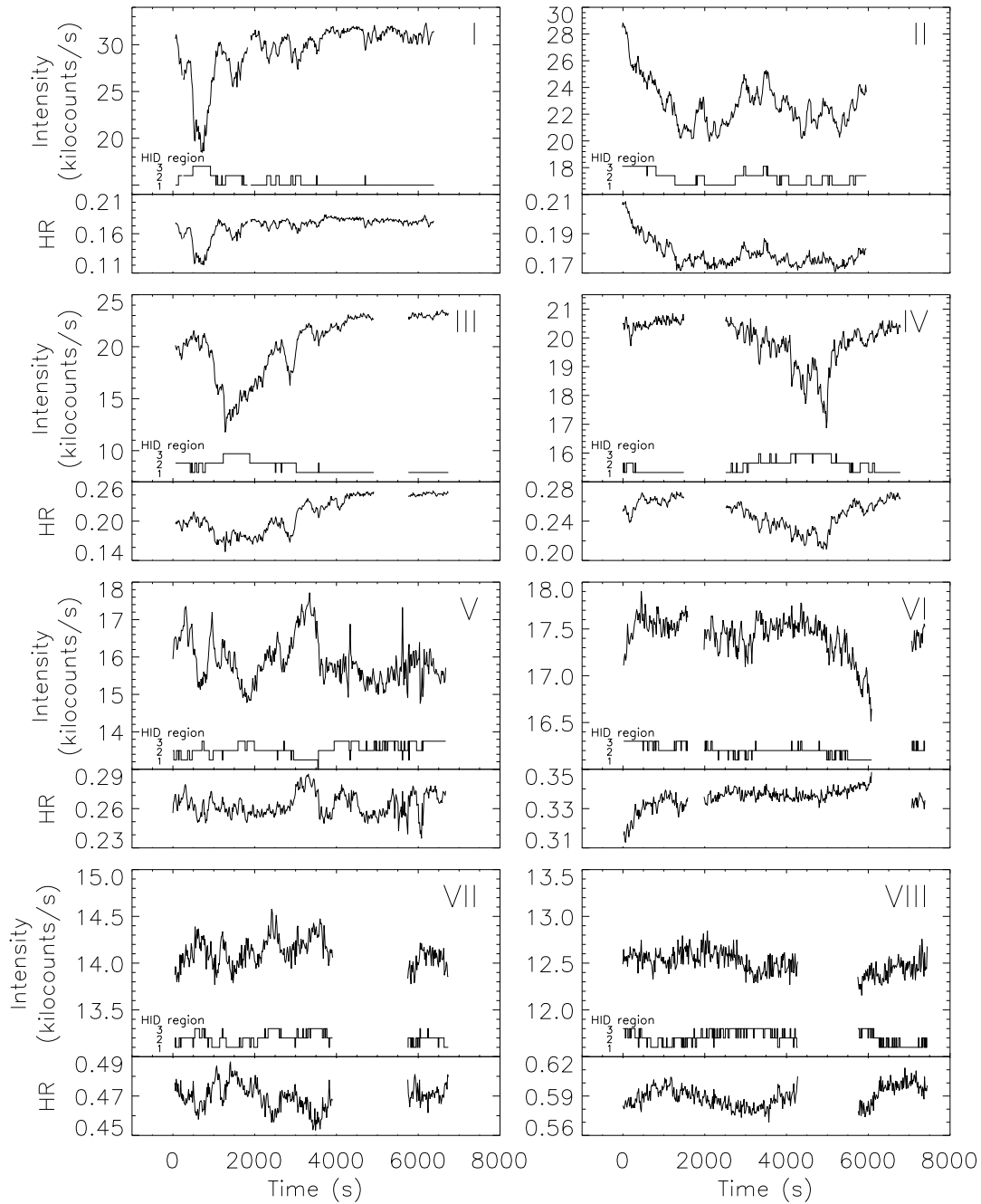}
\caption{ 
PCA light curves (2--18~keV, 5~PCUs) and hardness ratios (HR =
\broadcolor) in 16~s time bins for the eight observations made during
1997 February~18 -- March~4. A count rate of 13~kcts/s $\approx$
1.0~Crab.  The data gaps in Obs.~VI were longer than as shown here;
the second segment of the observation has been shifted left by 4000~s
and the third segment by 5000~s. These data were used to construct the
hardness-intensity diagram in
Figure~\protect{\ref{fig:cchid_all}}. The association with specific
regions of that diagram is indicated below each light curve.
\label{fig:pcalc}}
\end{figure*}

Figure~\ref{fig:pcalc} shows the light curves and hardness ratios
(with 16~s time resolution) for each of the PCA observations (I--VIII
in time order), made as the intensity declined from 2.5~Crab to
1.0~Crab.  On time-scales of hundreds of seconds, the observations
made at high intensities show strong variability, while observations
at 1.0~Crab show quite steady count rates. As expected from the ASM
hardness ratios, the PCA hardness ratio gradually increases from a low
value during the early observations when the source was in the flaring
state to a factor four higher as it reached the quiescent level. The
relationship between intensity and spectral changes is discussed in
detail below.

\section{Analysis and Results}

\subsection{Color-color and Hardness-intensity Diagrams}

\begin{figure}
\epsscale{1.25}
\plotone{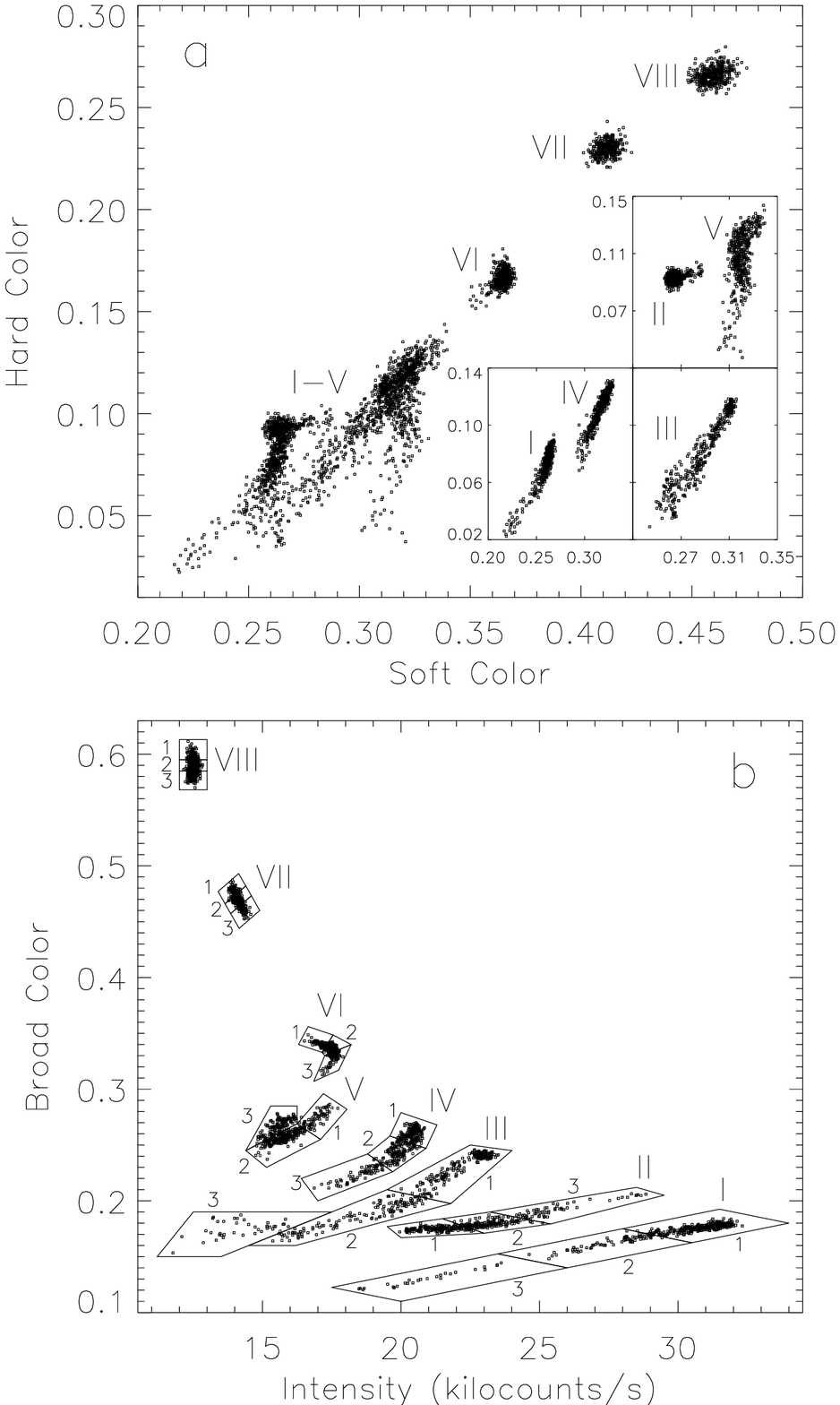}
\caption{ 
Color-color diagram (a) and hardness-intensity diagram (b) for all
eight observations (I--VIII). In the CD, soft color is defined as
\softcolor\ and hard color as \hardcolor\@. In the HID, the intensity,
\intens{2.0}{18}, is from all five PCUs and the hardness ratio is a
``broad'' color: \broadcolor\@. Each point corresponds to 16~s of
data. Background has been subtracted, but it does not affect the
intensity or soft color and only slightly affects the hard color. The
three insets in the CD separate overlapping points from observations
I--V. The HID track for each observation has been divided into three
regions (1--3) for timing analysis.
\label{fig:cchid_all}}
\end{figure}

For the eight PCA observations of 1997 February--March, 16~s intensity
and hardness-ratio measurements were used to construct color-color and
hardness-intensity diagrams (CDs/HIDs, Figure~\ref{fig:cchid_all}).
The hardness ratios were defined as the ratio of count rates in
selected energy bands: a soft color (\softcolor) and hard color
(\hardcolor) for the CD, and a broad color (\broadcolor) for the HID.
The evolution from flaring to quiescent state produced a large range
of colors and intensities over the entire cycle.  In contrast, each
individual observation yielded a localized cluster or track within the
CDs and HIDs. The spectral branches for each of the observations are
easier to distinguish in the HID than the CD.  The long tracks in the
HID associated with observations~I--V show the color changes
associated with the large intensity variations during the flaring
phases.  The intensity variations are smaller for observations
VI--VIII, but significant color changes do occur during these
observations as well.

The choice of energy bands used in constructing these diagrams can
affect the appearance of spectral tracks. For observations showing a
single branch, only the length and slope of the branch is affected.
Observations V and VI each show two branches. The orientation of these
branches is discussed in more detail below.

The tracks in the CD and HID are reminiscent of the correlated
spectral/intensity behavior of Z and atoll class LMXBs, which also
show correlations of temporal properties with position along tracks or
branches in CDs and HIDs (\cite{hk89}). Thus, we have investigated how
the temporal properties of Cir~X-1 are related to position in the CD
or HID. For this purpose, we divided the HID track for each of the
eight observations into three regions (Figure~\ref{fig:cchid_all}b).
The choice of numbers for each region was motivated in part by the
timing results discussed below, but the numbers serve mainly as
reference labels rather than as meaningful quantities (such as Z "rank
number").

\subsection{Power Density Spectra}

Fourier power density spectra (PDSs) were computed using 16 s segments
with 244~$\mu$s ($2^{-12}$~s) time bins. This was done for both the
full 2--32~keV energy range and for four energy channels: 2.0--4.8~keV,
4.8--13~keV, 13--18~keV, and 18--32~keV. The Leahy-normalized power
spectra (\cite{leahy83}) were converted to the fractional rms
normalization by dividing by the background-subtracted count rate in
the selected band. The expected Poisson level, i.e. the level of white
noise due to counting statistics, was estimated taking into account
the effects of deadtime (Morgan, Remillard, \& Greiner
1997\nocite{morgan97}; Zhang et~al.\ 1995\nocite{zhang95},
1996\nocite{zhang96}) and subtracted from each PDS; this method tends
to slightly underestimate the actual Poisson level. For each of the 24
HID regions defined in Figure~\ref{fig:cchid_all}, an average PDS was
calculated from the power spectra corresponding to points in that
region. The PDSs were then logarithmically rebinned.

\begin{figure}
\epsscale{1}
\plotone{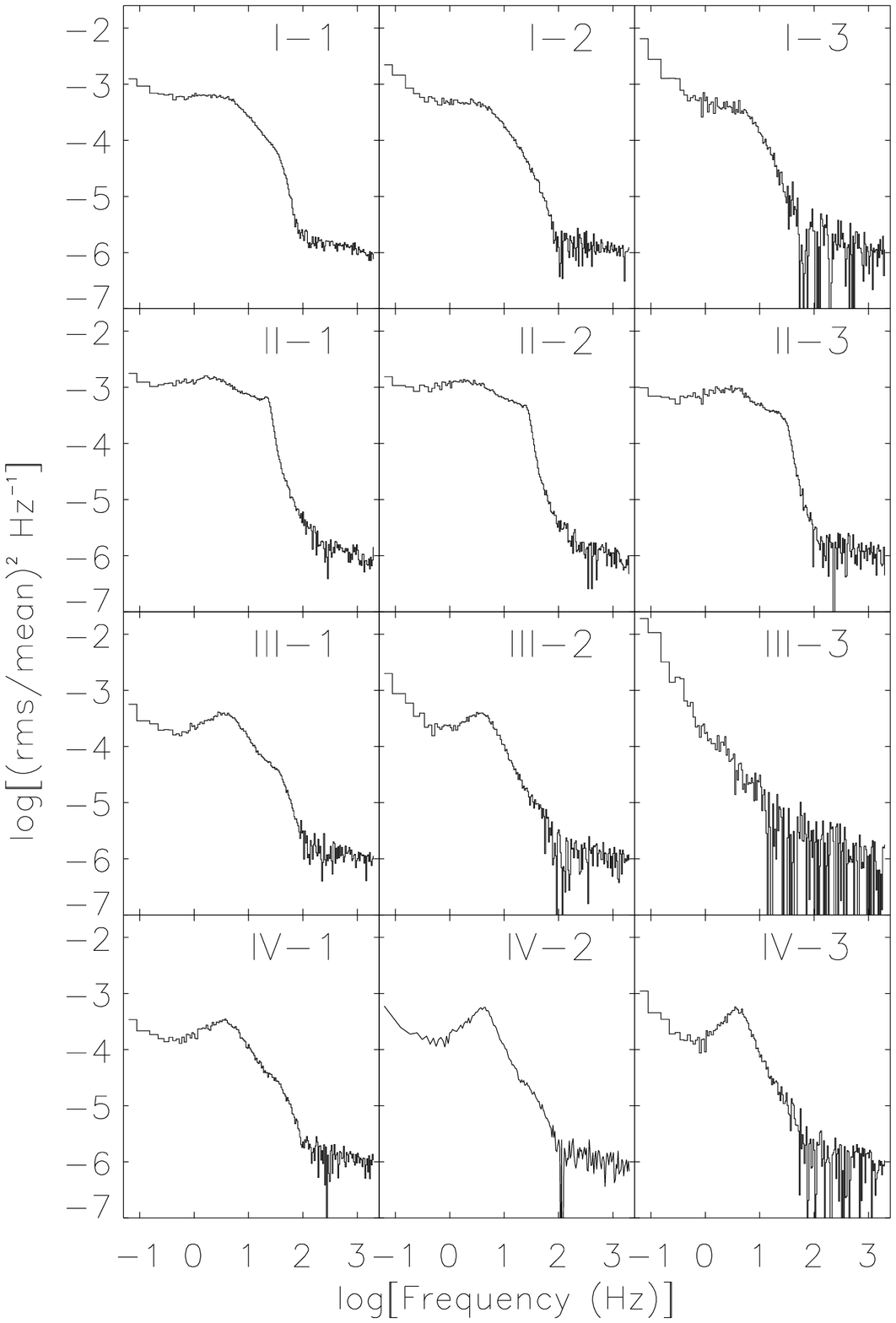}
\newpage
\plotone{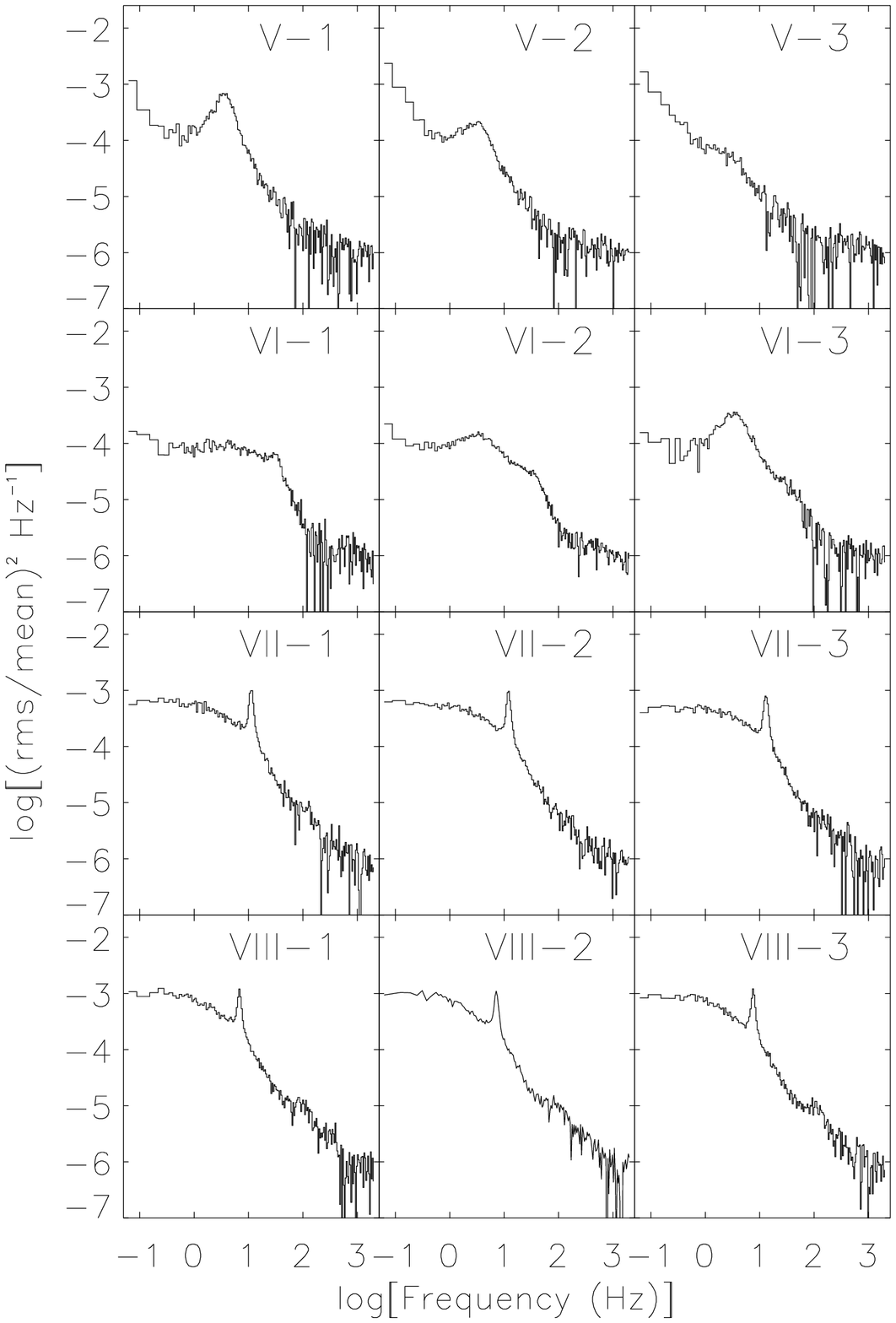}
\caption{ 
Averaged and rebinned power density spectra (2--32~keV) for each of
the three HID regions for each observation. Poisson noise has been
subtracted from each PDS (see text).
\label{fig:pds}}
\end{figure}

\begin{figure}
\epsscale{1.7}
\plotone{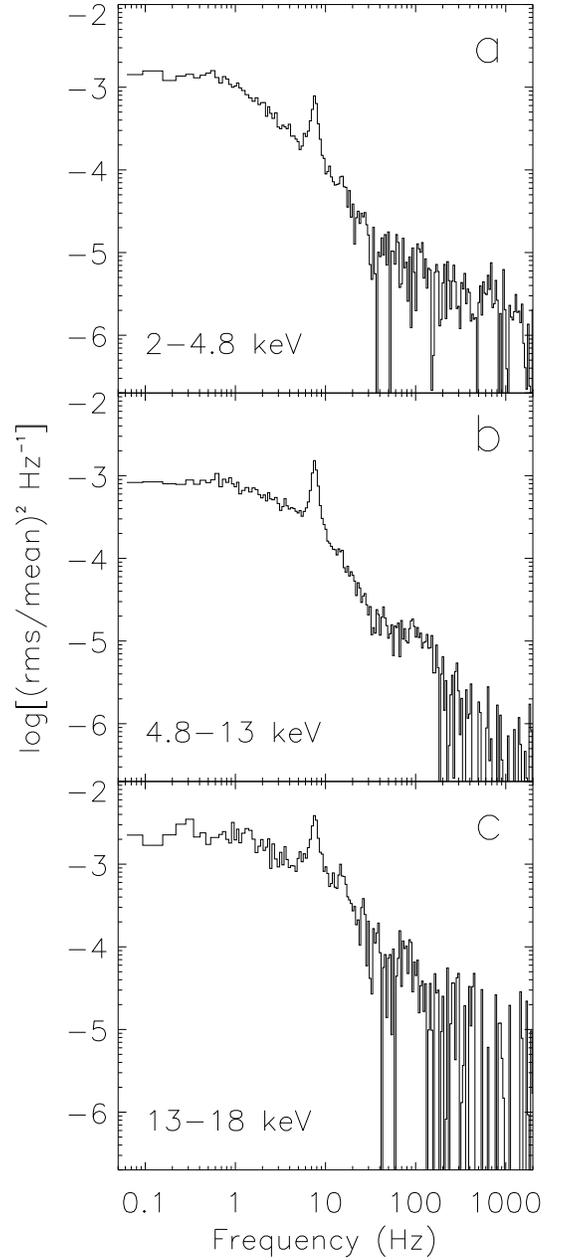}
\caption{
Averaged and rebinned power density spectra for HID region VIII-3 in
three energy bands. A harmonic peak of the 7.6~Hz QPO is clearly
visible in the high-energy channel (c).  The broad high-frequency
peak, most clear in (b), occurs near $\sim$100~Hz in this
observation. The low-frequency noise cuts off less sharply as energy
increases.
\label{fig:pds3E}}
\end{figure}

The average PDS (2--32 keV) for each HID region is shown in
Figure~\ref{fig:pds}.  During the extended active state (observations
I-VI), a broad peak is often observed near 4~Hz; this feature is
prominent in PDSs from observations III--VI, weak in observation~II,
and indistinguishable from a flat-topped component in observation~I
(see below). A strong narrow QPO feature is seen at frequencies from
6.8 to 13~Hz in observations VII and VIII.  In some cases, especially
at higher photon energy (see Figure~\ref{fig:pds3E}), a harmonic peak
is observed at twice the frequency of this QPO\@. A weak narrow QPO
feature is present at frequencies above 20 Hz in regions II-1 and
VI-1.  A sharp ``knee'' is present at similar frequencies in regions
II-2, II-3, III-1, IV-1, VI-2, VI-3, and possibly I-1. Broad
high-frequency noise is sometimes seen, e.g.\ at $\sim$100~Hz in
observation VIII (Figure~\ref{fig:pds3E}b). There is an underlying red
continuum spectrum of noise in all of the regions of the HID, but the
shape and low frequency slope of the continuum vary over a wide range.

The narrow QPO peaks and the low frequency noise in the PDSs from the
``quiescent'' 1~Crab observations (VII and VIII) resemble previously
observed PDSs (\cite{shirey96}; \cite{oosterbroek95};
\cite{tennant87}). Those PDSs also contained narrow QPO features, with
centroid frequencies in the range 1.3--20~Hz, and similarly shaped
low-frequency noise. The broad high-frequency component that we detect
in the present observations is similar to the 20--100~Hz QPO seen in
earlier PCA observations (\cite{shirey96}) and to the 100--200~Hz QPO
observed with \EXOSAT\ (\cite{tennant87}).

The weak narrow QPO feature above 20--30~Hz in regions II-1 and VI-1
occurs near the knee of the low-frequency noise component. This
similarity to the LFN and prominent QPO at lower frequency in
observations VII and VIII suggests these higher frequency oscillations
are produced by the same physical process as the lower frequency QPOs.
In observations II and VI, this QPO feature is visible in region 1 as
a small peak that fades in region 2 and becomes only a ``knee'' in
region 3 (see Figure~\ref{fig:pds}). Thus we assume that this knee is
related to the QPO.  Similar knees are present in regions III-1, IV-1,
and possibly I-1. We include a narrow QPO component in fits of PDSs
which show a knee above 20~Hz, but identify these cases as
``unpeaked'' in the discussion below.

Likewise, although no peak appears in the PDSs from observation I, a
broad noise component has roughly constant power below about 4~Hz and
drops off above that frequency, forming a ``knee'' which might
indicate the presence of the 4~Hz QPO component. The PDS for region
I-1 somewhat resembles those of regions III-1 and IV-1, in that all
show a break in the power spectrum near 4~Hz and a second knee or
change in slope near 30~Hz. We include a broad QPO component in fits
of the PDSs for observation I, but we identify these cases as
``unpeaked''.

The PDSs were fit with models comprising both broad-band and QPO
components: a power-law for the very low frequency noise (VLFN), an
exponentially cut-off power-law for the broad low-frequency noise, a
Lorentzian for the broad QPO near 4~Hz, Lorentzians for the narrow QPO
and its first harmonic, a broad Lorentzian for the high-frequency
peak, and a second power-law to fit the residual Poisson noise at high
frequency. The model for each PDS consisted of two to five of these
components, depending on which components were necessary for an
acceptable fit. The frequency of the harmonic (when present) of the
narrow QPO was fixed at twice the fundamental frequency.  For the fits
of the PDSs from the four narrower energy channels, the QPO centroid
frequencies were fixed at the values determined from the 2--32~keV
PDSs.  There were generally not enough counts to obtain useful PDS
fits for the 18--32~keV channel.  For use in performing the fits, we
estimated the variance of each power in each binned and averaged PDS
by calculating the sample variance of the powers in the individual
PDSs that were averaged to obtain each point, and dividing the result
by the number of the powers used in computing the sample variance.

The centroids of the narrow variable-frequency QPO and the $\sim$4~Hz
QPO were measured accurately whenever a clear peak was
visible. However, in cases where these components are weak or
unpeaked, the centroids were less well-constrained. The centroid of
the broad high-frequency peak and the cut-off frequency of the LFN
were often poorly constrained.

\begin{figure}
\epsscale{1}
\plotone{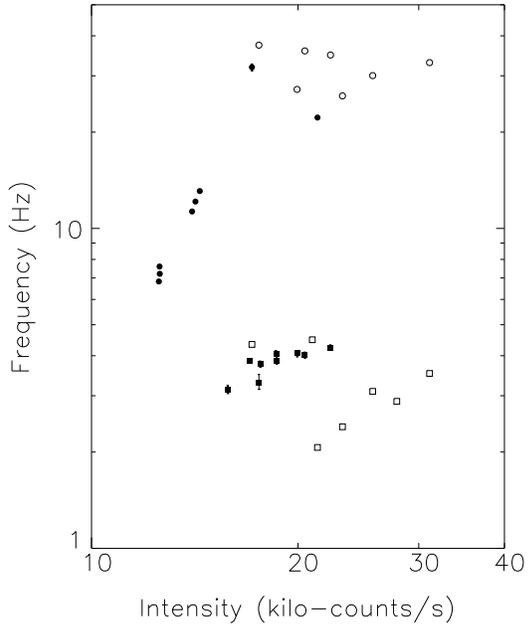}
\caption{ 
Centroid frequency of the QPOs versus intensity (\intens{2.0}{18})\@.
A filled circle represents the narrow QPO and a filled square
represents the broad $\sim$4~Hz QPO (all points below 5~Hz are the
broad QPO\@). Unfilled circles and squares indicate the approximate
frequency of a knee or very weak peak that may be associated with the
narrow and broad QPO respectively. Error bars on frequency
measurements (filled points only) represent 90\% confidence intervals
for a single parameter ($\Delta\chi^2$=2.7).  In many cases, the error
bar for the QPO frequency is smaller than the plot symbol.
\label{fig:freq_vs_I}}
\end{figure}

Figure~\ref{fig:freq_vs_I} shows the frequency of the broad and narrow
QPOs versus intensity (2--18~keV). The frequency of the narrow peak is
generally correlated with intensity, starting at 6.8~Hz at 1~Crab and
reaching 32~Hz at 1.3~Crab. At higher intensity, this QPO is sometimes
present above 20~Hz and is often unpeaked (i.e., a knee). In
observations III--VI, the broad QPO is clearly present at 3.3 to
4.3~Hz. This QPO component was included in the fits of the PDSs from
observations~I and II, and the resulting frequencies (2.1--4.5~Hz) are
shown as unfilled squares (indicating a weak peak or a knee) above
20~kcts/s in Figure~\ref{fig:freq_vs_I}.

The ratio of the width of the narrow QPO peak to its centroid
frequency ($\Delta\nu/\nu$) is about 0.15 when at 6.8 to 13~Hz. At
higher frequency this QPO becomes broader, with $\Delta\nu/\nu \sim
0.4$. When the broad QPO near 4~Hz is strong, we find that
$\Delta\nu/\nu \sim 1$, and when it is weak $\Delta\nu/\nu \sim 2$ to
$3$.

\begin{figure}
\epsscale{1.1}
\plotone{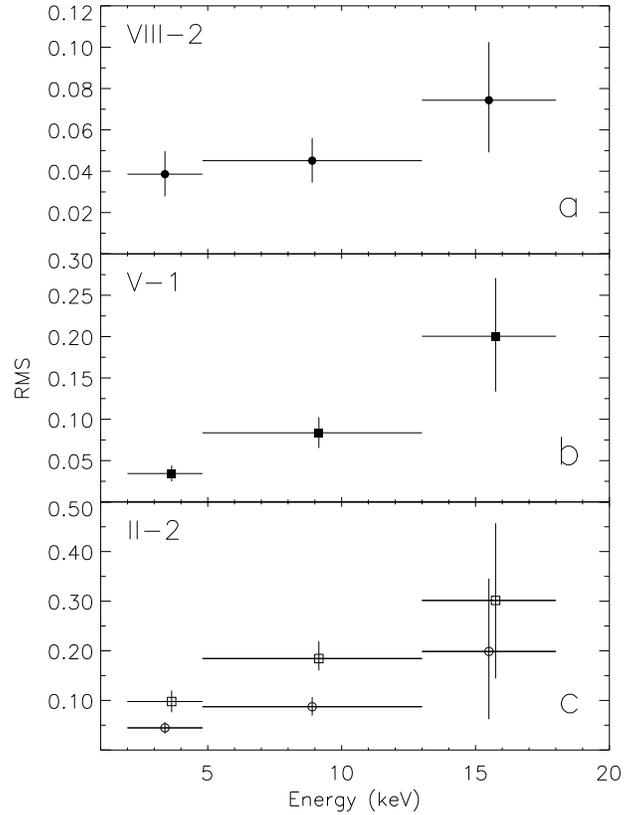}
\caption{ 
Rms amplitude of QPOs versus photon energy. Panel (a) shows
the rms for the narrow QPO at 7.2~Hz (solid dot) and (b) for an example
of the broad 4~Hz QPO (solid box). In panel (c), unfilled circles and
boxes indicate the rms amplitude of a component forming a knee or very
weak peak that may be associated with the narrow and broad QPOs
respectively. The broad QPO points have been offset slightly to the
right in energy for clarity. Errors on QPO amplitudes represent 90\%
confidence intervals.
\label{fig:rms_vs_E}}
\end{figure}

Figure~\ref{fig:rms_vs_E} illustrates the dependence of the rms
amplitude of the QPOs upon photon energy. Typical values for the rms
amplitude of the 6.8--13.1~Hz QPO at 2--4.8~keV, 4.8--13~keV, and
13--18~keV are 4\%, 5\%, and 8\% respectively
(Figure~\ref{fig:rms_vs_E}a), indicating a weak trend of increasing
rms amplitude at higher photon energy.  The amplitude of the broad QPO
increases significantly at higher photon energy. For clearly peaked
4~Hz QPOs, the rms amplitude is typically about 3\%, 8\%, and 18\% in
these three energy bands (Figure~\ref{fig:rms_vs_E}b). The rms values
vary considerably when these components are weak or unpeaked but their
amplitudes still generally increase with energy
(Figure~\ref{fig:rms_vs_E}c).

\subsection{Temporal Behavior versus Position on Spectral Branches}

\begin{figure}
\epsscale{1.1} 
\plotone{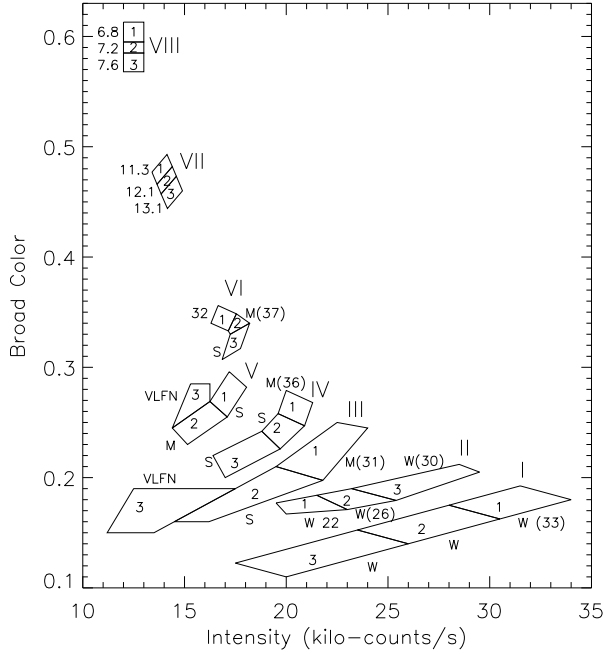}
\caption{ 
Hardness-intensity diagram showing QPO properties for the
regions from Figure~\protect{\ref{fig:cchid_all}}b. The frequency of
the 6.8 to 32~Hz QPO is labeled (in Hz) beside each region where it is
present. Parenthesized frequencies indicate that this component was
unpeaked, i.e., a knee. Letters indicate the strength of the broad
4~Hz QPO: S---strong, M---medium, and W---weak or unpeaked.
\label{fig:hidqpo}}
\end{figure}

The outlines of the HID regions of Figure~\ref{fig:cchid_all}b are
reproduced in Figure~\ref{fig:hidqpo} with labels summarizing the
observed QPO properties.
As the hardness ratio decreases and the intensity increases along the
HID tracks for observations VIII, VII, and VI, the frequency of the
narrow QPO feature increases from 6.8~Hz to 32~Hz. The feature is
rather weak and knee-like in observation VI, but it appears to have a
width consistent with the width of the prominent QPO peak in
observations VII and VIII. A similar weak and somewhat knee-like
feature is also present in observation II, where it increases in
frequency from 22~Hz to 30~Hz as the intensity increases. The PDSs
from observations I, III, and IV all show a knee above 30~Hz at the
high-intensity, hard end of their HID tracks; these knees may be
related to the narrow QPO features seen in the other observations.

The broad 4~Hz QPO is not present in the ``quiescent'' observations
(VII and VIII). This QPO is strongest in portions of the
intermediate-intensity observations (III--VI) and is weakly present in
the soft, high-intensity observations (II and possibly I).

Very low frequency noise dominates the power spectrum of regions V-3
and III-3. Both of these regions appear to begin upturned branches at
the low-intensity, soft end of branches showing the more pronounced
4~Hz QPOs.

\section{Discussion}

The combined temporal and spectral-branch properties of the
observations presented here suggest Z-like behavior. We identify the
6.8--32~Hz QPOs as horizontal-branch oscillations (HBOs), the 4~Hz QPO
as normal-branch oscillations (NBOs), and the strong VLFN as
flaring-branch behavior (see discussion below).  These identifications
of characteristic time-variability patterns then help to identify the
tracks in the HID as horizontal, normal, and flaring branches
(HB/NB/FB), where each 6 ks observation of Cir~X-1 appears to have
captured a snapshot of portions of one or two of the branches.  The
spectral branches appear to shift around as the flaring gradually
subsides, rather than forming a stable Z pattern. It is likely that
the shapes of the spectral branches become distorted somewhat during
these large shifts.  We now describe the inferred properties of each
of the spectral branches in more detail.

\subsection{Horizontal Branch}
HID regions VIII, VII, and VI-1 show a narrow QPO peak or knee at
6.8--7.6~Hz, 11.3--13.1~Hz, and 32~Hz respectively. This frequency
range overlaps the 13--60~Hz range of typical horizontal branch QPOs
(\cite{klis95}). The associated low-frequency noise and harmonic peak
are also typical of horizontal branch power spectra. The broad high
frequency peak in Cir~X-1 may be related to the high frequency noise
component often observed on the horizontal branch.

The HID track for observation VI shows the narrow QPO at 32~Hz on a
roughly horizontal segment (region VI-1) and a knee at 37~Hz on the
right end of this segment (region VI-2). The apex of region VI-2
brings a transition to the 4~Hz QPO, which is dominant on the downward
branch of this track (region VI-3). This is very similar to the HB/NB
transition in Z sources. 

When Cir~X-1 is in ``quiescence'' in observations VII and VIII, the
``horizontal branch'' turns upward and becomes vertical in the HID.
For comparison, \RXTE\ PCA observations of Cir~X-1 from 1996 March
10--19 which show a narrow QPO peak at 1.3--12~Hz (\cite{shirey96})
are almost entirely confined to the 12.3--14.7 kcts/s (2--21 keV)
intensity range. The HID tracks for those observations lie along a
nearly vertical line, and probably represent sections of the
``horizontal'' branch.

Observation II may also be on part of the HB, since a weak narrow QPO
appears to evolve into a knee and increase in frequency from 22 to
30~Hz as the intensity increases. However, the broad QPO is also
weakly visible in PDSs for this observation. The fact that
observations II, VI, VII, and VIII all show little variation of the
hard color used in Figure~\ref{fig:cchid_all}a suggests that
observation~II may be associated with the other HB observations.

\subsection{Normal Branch}

The 4~Hz QPO is observed when the source intensity rises above the
``quiescent'' 1-Crab level ($\sim$13~kcts/s). It is roughly stationary
in frequency (3.3--4.3~Hz when clearly peaked) and broader than the
HBO. The feature is easily seen in observations III--VI; at these
times the location in the HID moves along diagonal tracks. The
$\sim$4~Hz frequency and motion along diagonal tracks in the HID is
consistent with the 4--7~Hz NBOs observed at nearly constant frequency
on the NB of typical Z sources (\cite{hk89}). We therefore identify
the broad 4~Hz QPO as a normal branch oscillation, and the diagonal
tracks for observations III--VI as shifted normal branches.

The broad QPO component may be also present in the highest intensity
observations, as a weak feature in observation II and in the form of a
break near the 4~Hz QPO frequency in observation~I.  We also note that
at the top of the normal branch (regions I-1, III-1, IV-1, VI-2) a
knee above 30~Hz is present in addition to the NBO component.

A similar broad 4~Hz QPO is present in observations from 1996 March
5--6 made immediately before phase zero of the cycle showing the
1.3--12~Hz narrow QPO.

\subsection{Flaring Branch}
Beyond the left apex of the normal branch a short upturned branch is
observed in HID region~V-3 and possibly III-3. The PDS for these
regions are dominated by very low frequency noise, which is typical
for flaring branches, and no QPO peaks are obviously apparent. We note
that in the well-established Z sources neither NBOs nor HBOs are
present on the flaring branch, except for Sco~X-1 and GX~17+2, in
which the NBO evolves into a 6--20~Hz QPO (\cite{klis95} and
references therein).

\begin{figure}
\epsscale{1}
\plotone{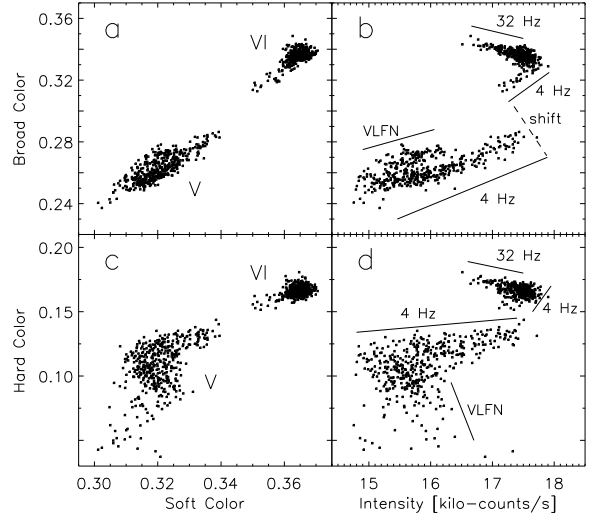}
\caption{ 
Broad-color and hard-color CDs and HIDs for observations~V and VI.  In
all four diagrams, Obs.~V is in the lower left and Obs.~VI in the
upper right. The CD and HID tracks for Obs.~V both turn upward at the
left end in the broad-color diagrams (a,b) but turn downward in the
hard-color diagrams (c,d). In the HIDs, presence of the 32~Hz HBO,
4~Hz NBO, and VLFN is indicated along the branches. An apparent shift
of the normal branch between observations~V and VI is labeled in (b).
The intensity, \intens{2.0}{18}, is from all five PCUs. The soft color
is defined as \softcolor, the broad color as \broadcolor, and the hard
color as \hardcolor\@. Each point corresponds to 16~s of
background-subtracted data.
\label{fig:cchid_56medhard}}
\end{figure}

The left end of the spectral track for observation~V bends upward in
the HID shown in Figure~\ref{fig:cchid_all}b, but bends downward in
the CD in Figure~\ref{fig:cchid_all}a. This behavior is demonstrated
more clearly in Figure~\ref{fig:cchid_56medhard}, which shows CDs and
HIDs for observations V and VI. When a broad color (\broadcolor) is
used as the ordinate of the diagrams
(Figure~\ref{fig:cchid_56medhard}a,b), the track for observation~V
turns upward on the left end. When a harder color (\hardcolor) is used
as the ordinate (Figure~\ref{fig:cchid_56medhard}c,d), this branch
turns downward. The CD and particularly the HID version based on the
harder color show the most clear similarity to canonical Z diagrams,
with the temporal behavior of observations V and VI being generally
consistent with horizontal, normal, and flaring branches. The
broad-color HID (Figure~\ref{fig:cchid_56medhard}b) shows evidence for
a shift of the normal branch that does not show up in the other three
diagrams of that figure.

\subsection{Relation to Other Sources}
Our observations reveal spectral branches which shift in the CD and
HID as Cir~X-1 evolves from a soft, high-intensity state to a hard,
lower-intensity state. The ASM light curves and hardness ratios
(Figure~\ref{fig:asm}) show that this evolution occurs periodically
with the 16.55~day cycle, thus suggesting that the CD/HID shifts may
also be periodic. Shifts of the ``Z'' pattern in CDs and HIDs have
been observed in the so-called Cyg-like Z sources: Cyg~X-2 (Kuulkers,
van~der~Klis, \& Vaughan 1996\nocite{kuulkers96:cygx-2}), GX~5-1
(\cite{kuulkers94:gx5-1}), and GX~340+0
(\cite{kuulkers96:gx340+0}). However, the shifts do not occur
periodically in those sources, nor do they have the magnitude of the
shifts observed in Cir~X-1.

The flaring branch of Cir X-1 turns upward when a soft or broad color
is used on the vertical axis. When a harder color is used, this branch
turns downward but then bends to the left. In the Cyg-like Z sources,
the flaring branch sometimes turns upward or starts toward higher
intensity and then loops back to lower intensity
(\cite{kuulkers96:cygx-2}; \cite{kuulkers94:gx5-1};
\cite{kuulkers96:gx340+0}; \cite{penninx91:gx340+0}). In some cases,
these sources are observed to ``dip'' while on the flaring branch
(\cite{kuulkers94:gx5-1}; \cite{penninx91:gx340+0};
\cite{wijnands97}), with tracks which turn down and then to the
left, similar to that of Cir~X-1 in Figure~\ref{fig:cchid_56medhard}c.

The left end of the horizontal branch in Cir~X-1 turns upward and
becomes vertical at low intensity (Figure~\ref{fig:hidqpo}). On this
section of the branch, HBO frequencies are low: 6.8--13~Hz in
observations VII and VIII and 1.3--12~Hz in the earlier 1996 March
observations. A similar effect was reported in GX 5-1
(\cite{lewin92:gx5-1}; \cite{kuulkers94:gx5-1}), in which the HB turns
upward at the low-intensity end while HBOs are observed at relatively
low frequency (13--17~Hz). In fact, Lewin et~al.\
(1992\nocite{lewin92:gx5-1}) suggested that other Z sources might show
such an upward turn of the HB if their intensities and QPO frequencies
became sufficiently low.

The 5--20~Hz narrow QPO was detected with \EXOSAT\ at an intensity
similar to the quiescent level observed by \RXTE. We note that
absorption dips are responsible for much of the structure seen in the
CD shown for that observation; however, the HIDs show that the narrow
QPO occurred on an upturned left end of a horizontally oriented track
as in our data (see Figures~2--4, 8, \& 10 in \cite{oosterbroek95}).
At higher intensity during the same observation, the narrow QPO was
not present, and we note that some of the high-intensity PDSs show
hints of a broad peak near 4~Hz. We thus conclude that the behavior
observed by \EXOSAT\ during that observation is related to the Z-like
behavior we observe with \RXTE.

Most of the other \EXOSAT\ observations took place when Cir~X-1 was
significantly lower in intensity than the ``quiescent'' level of the
current observations. The CDs and HIDs for these \EXOSAT\ observations
did not show tracks which could clearly be identified as Z or atoll.
Their power spectra were generally dominated by VLFN, typical of atoll
sources in the banana state, and sometimes also showed a broad red
noise component resembling atoll high-frequency noise
(\cite{oosterbroek95}). However, these power-spectral shapes are not
unique to atoll sources: power spectra for black hole candidates in
the high state are dominated by VLFN, as are those of the current
observations on the low-intensity end of the normal branch and on the
flaring branch (i.e., regions III-3 and V-3).

Cir~X-1 was expected to never show HBOs since atoll-like behavior was
taken as evidence that the magnetic field is not strong enough to
allow the magnetospheric beat frequency mechanism (MBFM) to operate
(\cite{klis94}; \cite{oosterbroek95}). (However, it is also possible
that the HBOs are not produced by the MBFM\@.) The results presented
here demonstrate both HBOs and NBOs in Cir~X-1 and show no evidence
for atoll behavior. Since the atoll-like behavior observed with
\EXOSAT\ occurred at lower intensity than in the present observations,
it is possible that they do represent a different state of the
source. If Cir~X-1 actually can show atoll behavior as well as the
Z-like behavior shown here, then we would have new clues to the
differences between the two types of sources. Such observations would
challenge the hypothesis that differences in both $\dot{M}$ and
magnetic field distinguish these two classes.

\section{Summary}
Our results from an analysis of \RXTE\ observations of Cir~X-1 reveal
behavior similar to that of Z sources, and, in particular, allow us to
identify temporal and spectral signatures of the horizontal, normal,
and flaring branches.  The spectral variability of Cir~X-1 is seen to
correspond to tracks in a HID which are similar in direction to the
typical direction in the HID of Z sources in general, but the
locations of the tracks corresponding to each branch move from
observation to observation in a systematic manner. 

To be specific, in the current observations of Cir~X-1 the horizontal
branch is characterized by the presence of relatively narrow
6.8--32~Hz QPO features in the PDS. The track in the HID of the
horizontal branch is horizontal at the high intensity end and becomes
vertical at the low intensity end, where the source is ``quiescent'',
i.e., has an intensity near 1~Crab and is characterized by a
relatively low degree of variability on time scales longer than 1~s.
The normal branch is characterized by broad 4~Hz QPOs, and by motion
in the HID which generally falls along tracks which run diagonally
from hard high-intensity locations to soft low-intensity locations.
There are also time intervals when the PDS is dominated by very low
frequency noise. We identify these intervals as excursions onto the
flaring branch.

The large amplitude intensity variations associated with the
active/flaring state of Cir~X-1 can be divided into three categories:
(1) motion across the horizontal portion of the horizontal branch and
along the normal and flaring branches, (2) shifts of the spectral
branches, and (3) absorption dips.  While our \RXTE\ observations have
allowed us to recognize and distinguish these different types of
variability, there is still much to be understood about the physical
mechanisms responsible.

\acknowledgments

We are grateful to the entire \RXTE\ team at MIT for their support. We
thank particularly R. Remillard for helpful discussions. Support for
this work was provided by NASA Contract NAS5-30612.

\newpage



\newpage

\newpage

\newpage

\newpage

\newpage

\newpage

\newpage


\newpage

\newpage


\clearpage
\begin{deluxetable}{cccc}
\footnotesize
\tablewidth{0pt}
\tablecaption{PCA observations of Cir~X-1 during 1997 February~18 -- March~4
\label{tab:obs}}
\tablehead{
\colhead{Obs.} & \colhead{Julian Date\tablenotemark{a}} & \colhead{Phase} &
\colhead{Mean Intensity\tablenotemark{b}} \\
\colhead{} & \colhead{} & \colhead{} & \colhead{(Crab)}
}
\startdata
I & 2450497.90 & 0.10 & 2.3 \nl
II & 2450499.98 & 0.23 & 1.8 \nl    
III & 2450501.69 & 0.33 & 1.6 \nl   
IV & 2450503.66 & 0.45 & 1.5 \nl      
V & 2450505.80 & 0.58 & 1.2 \nl      
VI & 2450507.31 & 0.67 & 1.3 \nl      
VII & 2450509.35 & 0.79 & 1.1 \nl      
VIII & 2450511.62 & 0.93 & 1.0 \nl      
\enddata
\tablenotetext{a}
{Midpoint of 2-3~hr observation ($\sim$6~ksec of data per observation)}
\tablenotetext{b}
{1.0~Crab $\approx$ 13,000~counts/s (2--32~keV)}
\end{deluxetable}


\end{document}